\documentstyle[12pt,epsf]{article}
\def \a {\alpha}

\def \d {\partial}
\def \s {\sigma}

\def \e {\epsilon}
\def \t {\theta}
\def \T {\Theta}
\def \p {\phi}

\def \be {\begin{equation}}
\def \ee {\end{equation}}
\def \l {\lambda}
\def \w {\omega}
\begin{document}
\begin{titlepage}
\begin{flushright}
SU-4240-690\\
IMSc 98/07/40\\
\today
\end{flushright}

\centerline{\LARGE  Knot Solitons}

\begin{center}
{\bf T. R. Govindarajan$^\dagger$}\\
{\em Dept of Physics, Syracuse University\\
Syracuse, NY 13244-1130, USA}\\
and\\
{\em The Institute of Mathematical Sciences$^{\dagger\dagger}$\\
Chennai 600 113, India}\\
\end{center}
\vskip1cm
\noindent{\bf Abstract:} The existence of ring-like and knotted solitons
in O(3) non-linear $\s$ model is analysed. The role of isotopy of knots/links
in classifying such solitons is pointed out. Appearance of torus knot
solitons is seen.

\vfill
\hrule
\vskip1mm
\noindent{\em $\dagger$trgovind@suhep.phy.syr.edu; trg@imsc.ernet.in}
\noindent{\em $\dagger\dagger$ permanent address}
\vskip.5cm
\end{titlepage}

\noindent{\bf Introduction:}
Recently the possible existence of knotlike solitons in nonlinear 
field theories has been argued{\cite {FN1,FN2}. 
The toroidal solitons have been studied in 
3-d field theories for the past two decades{\cite {K,deVega,wu}.
Several numerical attempts  using various ansatzes have been made
to study these solitons{\cite {N1,N2}.
There also exists contradicting results regarding the ringlike nature 
of the solitons for low values of topological quantum numbers{\cite{N3}}.
The exciting possibilty of the Nonabelian gauge theories being
described by new variables which will have the component of nonlinear sigma
model have been pointed out{\cite{FN2}}. This will lead to presence
of exotic solitons and new applications in QCD. 

 In this letter we point out the arguments
in favor of knotted solitons and ways of chracterising them, in the specific 
well studied O(3) sigma model. We point out  that solitons in this 
model are characterised in addition to the Hopf number also by the genus 
of the seifert surface for the torus knot. We also point out only torus
knots appear in these models.

The O(3) nonlinear sigma model in 3 + 1 dimensions is defined by a vector 
field {\bf n} at every point in the space. The configuration 
space of the systemis characterised by maps from $R^3 \longrightarrow S^2$.
 The action can be written as 

\begin{equation} S~=~\int~d^4x ({\alpha^2} (\partial_\mu {\bf n})^2 + 
\epsilon^2({\bf n}.\partial_\mu {\bf n}~\times~ \partial_\nu {\bf n})^2~
+\nu (1~-~n_3))\label{action}
\end{equation}

This is the well known Faddeev model{\cite {F}}. 
The unit vector field {\bf n(x)}goes to a constant 
vector say $~col~(0,0,1)$ at $\infty$. The configuration space for such 
a model is then specified by maps from $S^3~\longrightarrow ~S^2$. 
Normally the solitons sectors for such a system are 
obtained  by the path-connectedness or $\Pi_0(configuration~space)$
{\cite {S,Fi,B}}. 
This in this case is given by 
\be
\Pi_0(Q)~=~\Pi_3(S^2)~=~{\cal Z}
\ee

Without the presence of the fourth order term in the action 
the solitons would prefer to be have 
zero size which can be seen by rescaling{\cite{FN1}}. 
Because of the second term the finite size solitons 
can exist. The second term is also unique in the sense it is the term that 
can be added with a maximum of two time derivatives required for second order 
equations of motion.The symmetry of the action is $SU(2)\otimes SU(2)$. 
It was shown in {\cite {K}} that the symmetry of the 
nontrivial soliton cannot be more than $~Diag~(U(1)\otimes U(1))$. 
It is easy to see how these solitons which are known as torons arise.
They are classified by the appropriate 
hopf index of the maps $S^3~\longrightarrow~S^2$. It is also 
known{\cite {K,K1}} that soliton mass obeys the bound given by 
\be
H~\geq~\epsilon~\alpha~(4\pi)^2~3^{3 \over 8}~{\sqrt 2}~|Q|^{3 \over 4}.
\ee
where Q is the topological charge. 

It has to be understood that the actual maps have to be differentiable 
in order to make sense in the Hamiltonian. Also we will assume 
that the maps are exponentially falling off which is ensured by
the mass term added at the end of the action {Eq.(\ref{action})}. 
Hence in principle we should seek classification of maps which are 
differentiable and exponentially decreasing. It was argued numerically 
that torons with such boundary conditions can exist by many authors. 
It was also argued that in addition to such torons
(which we may call as unknots following standard terminology of knot theory)
there can also be knotted solitons particularly 
for the higher topological charges. 

Now we will look at these knotted solitons from a new perspective 
emerging from the well known 
techniques used to establish spin-statistics theorem in 2+1 dimensions. 
Consider the same nonlinear sigma model in 2+1 dimensions 
specified by the action,

\be
S~=~\int d^3x~({1\over 2}\partial_\mu {\bf n}~\partial^\mu {\bf n}) + S_{Hopf}
\ee

Here the solitons are labelled by $\Pi_0(S^2\longrightarrow S^2)$ 
which is also ${\cal Z}$. A typical soliton map of topological number $m$ 
is given by{\cite{P,R}}

\be 
{\bf n} = \pmatrix {\sin {\t(r)}\cos m\p\cr \sin {\t}(r)\sin m{\p} \cr \cos {\t}(r)}
\ee

where $\t(r) = 0 $ at $ \infty$ and $\pi$ at $r=0$.

The spin and statistics of such a soliton is to be obtained through 
the fundamental group of the configuration space which is 
$\Pi_3(S^2)~=~{\cal Z}$. As mentioned earlier this is given by 
Hopf index of the map. One can add a term to the action 
with a coefficient ${\T}$
which reproduces the spin and exchange statistics{\cite{W}}.  
This is included in the action and explicitly it is given by

\be
S_{Hopf}~=~{\T \over 4\pi}\int~d^3x~\e^{\mu\nu\l}~A_\mu~F_{\nu\l}
\ee
where $F_{\mu\nu}~=~\d_\mu~A_{\nu}~-~\d_\nu~A_\mu$. In this case 
Hopf index one configuration will 
have spin $={\T\over 2\pi}$. A typical time dependent configuration can be 
presented as a `soliton' and an
`anti-soliton' created at time $t_0$ and annihilated at time $t_1$. 
In between $t_0$ and $t_1$ the soliton is rotated through 
an angle $2\pi$.  This can be pictured as follows.
$$
\setlength{\unitlength}{.8cm}
\begin{picture}(5,4.5)
\put(0,0){\line(1,0){4}}
\put(0,4){\line(1,0){4}}
\put(.5,.5){\line(1,0){3}}
\put(.5,3.5){\line(1,0){3}}
\put(0,0){\line(0,1){4}}
\put(.5,.5){\line(0,1){3}}
\put(4,0){\line(0,1){.5}}
\put(4,3.5){\line(0,1){.5}}
\multiput(3.5,.5)(0,1.5){2}{\line(1,3){.5}}
\multiput(3.8,1.1)(0,1.5){2}{\line(1,-3){.2}}
\multiput(3.5,3.5)(0,-1.5){2}{\line(1,-3){.2}}
\put(.25,3){\vector(0,-1){2}}
\put(4.2,1){\vector(0,1){2}}
\put(2.0,-.6){Fig.1.}
\end{picture}
$$
\vskip.5cm

Such a configuration has Hopf index = 1 as can be seen by the linking number
of the preimages of two points{\cite{Bo}}.
This accounts for the spin of the 2+1 d soliton. 
If we consider soliton-anti soliton pairs
created  at two different locations exchange and further
annihilation of the solitons give the configuration pictured in Fig.2.
$$
\setlength{\unitlength}{.8cm}
\begin{picture}(12,5)
\put(0,0){\line(1,0){4}}
\put(0,4.5){\line(1,0){4}}
\put(.5,.5){\line(1,0){3.5}}
\put(.5,4){\line(1,0){3.5}}
\put(0,0){\line(0,1){4.5}}
\put(.5,.5){\line(0,1){3.5}}
\put(8,0){\line(1,0){4}} 
\put(8,4.5){\line(1,0){4}} 
\put(8,.5){\line(1,0){3.5}}   
\put(8,4){\line(1,0){3.5}}  
\put(12,0){\line(0,1){4.5}} 
\put(11.5,.5){\line(0,1){3.5}}
\multiput(4,0)(0,.5){2}{\line(1,1){4}}
\multiput(4,4.5)(0,-.5){2}{\line(1,-1){1.7}}
\multiput(8,0)(0,.5){2}{\line(-1,1){1.7}}
\put(.25,3.5){\vector(0,-1){2.5}}
\put(11.75,3.5){\vector(0,-1){2.5}}
\put(4.5,.75){\vector(1,1){3}}
\put(5.5,-.6){Fig.2.}
\end{picture}
$$
\vskip.5cm
Note that Fig.1 and Fig.2 correspond to maps which are homotopically 
and isotopically are equivalent.
These configurations will serve typically as 3 dimensional torons. 
Higher topological number torons
are obtained through rotation through $2n\pi$ in Fig.1. 
It is obvious the maximally symmetric configurations will 
have a symmetry $~Diag~(U(1)\otimes U(1))$. The size and thickness of 
the toron will be fixed by the parameters of the 3 dimensional action. 
If we proceed to construct the Hopf-index 
3 configuration through the exchange process we will get the following 
configuration.

\vskip1cm
\begin{center}
\epsfxsize 2in
\hspace{.5cm}
\epsfbox{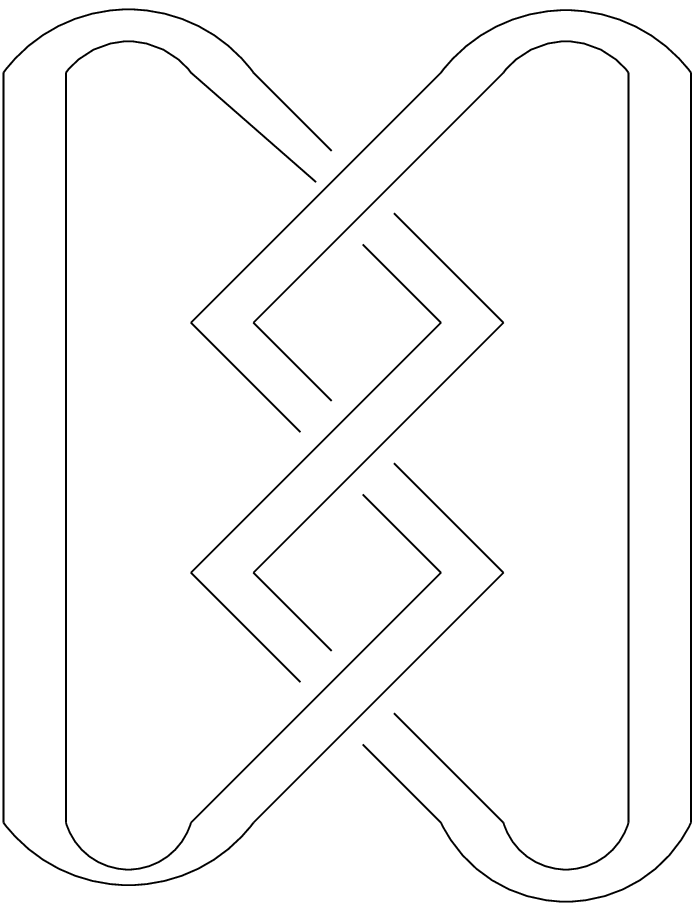}
\end{center}

It is easy to see that this is typically the trefoil knot soliton. 
Even though the Hopf-index is 3 and two configurations are homotopically
equivalent they are not `isotopic' 
to each other. {\em This brings in the question of isotopy of the links
of two preimages 
of the maps as an important tool to classify the solitons}.  
 
Now we shall focus on how to characterize the solitons in 3-dimensions. 
Though explicite solutions are not available these characterisations will
serve as a starting points of variational computations. This will be presented
in future paper. First we go to sterographic coordinates  to specify 
compactified $R^3 $($r^2~=~\sum x_i^2)$. 

\be 
X_i~=~{2x_i\over 1+r^2}~~~~i~=~1,2,3
\ee
and
\be
X_4~=~{1~-~r^2\over 1~+~r^2}
\ee
Define
\be 
Z_1~=~X_1~+~i~X_2,~~~~~~~~~~~Z_2~=~X_3~+~i~X_4
\ee
With these definitions it is easy to see $|Z_1|^2~+~|Z_2|^2~=~1$.
Similarly we can parametrize the unit vector field ${\bf n}$ by a 
complex function $\w$.
The standard definition is 
\be 
n_1~+~i~n_2~=~{2\w \over 1~+~|\w|^2}
~~~~~~~~~~~ and~~~~~~~~n_3~=~{1~-~|\w|^2\over 1~+~|\w|^2}
\ee

The action can be rewritten in terms of $\w$ as:

\be
S~=~\int{\a^2\over 4}{|d\w|^2\over (1+|\w|^2)^2}~+~
{\e^2\over 16}{(d\w\wedge d{\bar \w})^2\over (1+|\w|^2)^4}~+~\nu {2\over (1+|\w|^2)}
\ee

We can adopt the angular cordinates for $Z_1$ and $Z_2$,
\be Z_1~=~\cos \t e^{i\psi}~~~Z_2~=~ \sin \t~e^{i\chi}
\ee
Then $\w~=~\cot \t~e^{i(\psi~-~\chi)}$, will have Hopf number 1, but the 
energy density will not be ring like. A simple modification like 
\be
\w~=~f(\rho,z)~\cot \t~e^{i(\psi~-~\chi)}
\ee
with the function f falling off away from $\rho~=~1$ and $z~=~0$, will have the
requisite ring structure. The Hopf-index can be easily seen from the 
linking number{\cite{Bo}} of preimages of two points on $S^2$. The higher solitons
could be obtained from ${Z_1^m\over Z_2}$ or $\cot \t~e^{i(m\psi~-~\chi)}$.

To obtain the knotted solitons a known result from{\cite{mi}}. For example
if we consider 
\be g(Z_1,Z_2)~=~Z_1^2~+~Z_2^p \label{torus}
\ee
with p odd, then the surface
$g(Z_1,Z_2)~=~0$ cuts the $S^3$, namely $|Z_1|^2~+~|Z_2|^2~=~1$ along
a torus knot (2,p). When $p~=~1$ we get unknot. When $p~=~3$ we get the 
trefoil knot. So a map 
\be
\w~=~{Z_1^2~+~Z_2^3\over Z_2}
\ee will have the trefoil knot structure and Hopf-index~=~3. One should add
the damping function for variational minimisation of the energy. An
interesting offshoot is when p~=~even one obtains a link. For example
p~=~2 the map is
\be 
\w~=~{Z_1^2~+~Z_2^2\over Z_2}
\ee
and one gets Hopf link. This can be seen in the following exchange diagram.

\vskip.5cm
\begin{center}
\epsfxsize2in
\hspace{.5cm}
\epsfbox{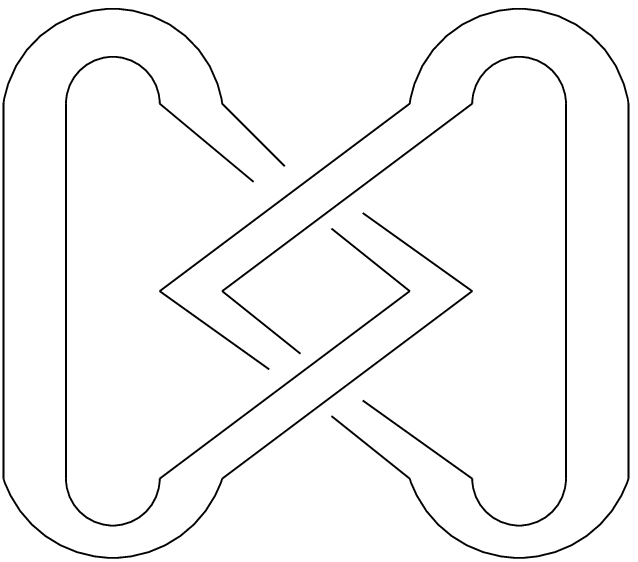}
\end{center}

This will lead to a ``molecule'' or
boundstate of two solitons of topological charge~=~1. Infact this kind of 
structures have been pointed out by Thompson himself while proposing
knots as model for atoms{\cite {T}}. The important characterisation for the 
`isotopy' of such knots would be the genus $g$ of the
Seifert surface bounded by the 
knot{\cite {mi}} which for the maps in {Eq.(\ref {torus})} $g~=~{p-1\over 2}$. 

In conclusion we have pointed out a new perspective to look at the 
ring like and knotted solitons in 3d starting from 2d solitons. Also a
new characterisation  is proposed on isotopy of maps which will be 
interesting tool to analyse these solitons.
This raises the important question that if such rings/knots are
the basic soiltons what is the statistics of such solitons.
Already some interesting work in this direction has been done{\cite {ba}}.
If such objects are present in the analysis of pure non-abelian gauge theories
as pointed out by Faddeev and Niemi it would be interesting to
quantize such solitons through effective actions and get the spectrum.
 Further work on variational
procedure based on these will be reported later.

\noindent{\bf Acknowledgements:} I would like to thank 
R. K. Kaul and A. P. Balachandran for discussions. This work had 
the support by the DOE through DE-FG02-85ER40231.

\end{document}